# The Distributed Computing Model Based on The Capabilities of The Internet

Lukasz Swierczewski
Computer Science and Automation Institute
College of Computer Science and Business Administration in Łomża
Lomza, Poland
luk.swierczewski@gmail.com

*Abstract*—**Paper describes the theoretical and practical aspects of the proposed model that uses distributed computing to a global network of Internet communication. Distributed computing are widely used in modern solutions such as research, where the requirement is very high processing power, which can not be placed in one centralized point. The presented solution is based on open technologies and computers to perform calculations provided mainly by Internet users who are volunteers.**

*Keywords-distributed computing, architectures and design systems*

## I. Introduction

According to the definition, distributed system is a collection of independent equipment connected together as a one seamless logical entity. Most commonly the term equipment is used to describe desktop computers, however, currently that can also include media tablets as well as handsets. The solution described in the article is based on the Internet, yet less sophisticated communication methods can also be used. Distributed systems create the illusion of a centralized system (single and integrated). This feature is called transparency and it is one of the key characteristic of this class of solutions. The concept of using distributed resources appeared in the late 70s. Today, thanks to the Internet, it is possible to exploit millions of computers provided by volunteers.

## II. General

There are various distributed system models. The most popular ones include: Peer-To-Peer, Cloud Computing, Grid Computing and Client/Server. In open-source solutions Client/Server architecture is most widespread. One of the perfect example of a Client/Server model is BOINC (Berkley Open Infrastructure for Network Computing), that started as a resource pooling solution for SETI@Home project and is constantly being developed at University of California, Berkeley.

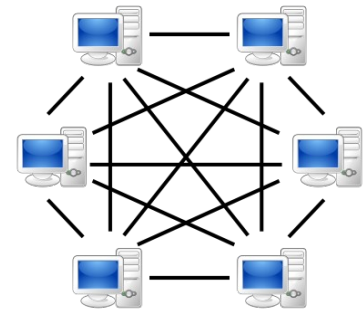

Figure 1. Peer-To-Peer Architecture Diagram

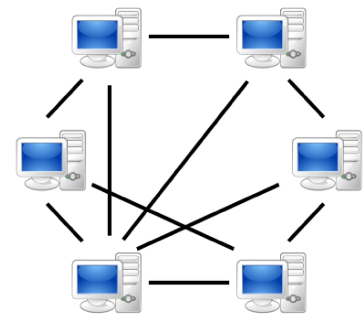

Figure 2. Client/Server Architecture Diagram

## III. Prepare Your Paper Before Styling

The principle behind BOINC is quite straightforward. Each Internet user can download the client application, that automatically connects to the server and downloads selected data portions for further computation. After the assigned task is completed, client uploads the results to the server where scientists can analyze them.

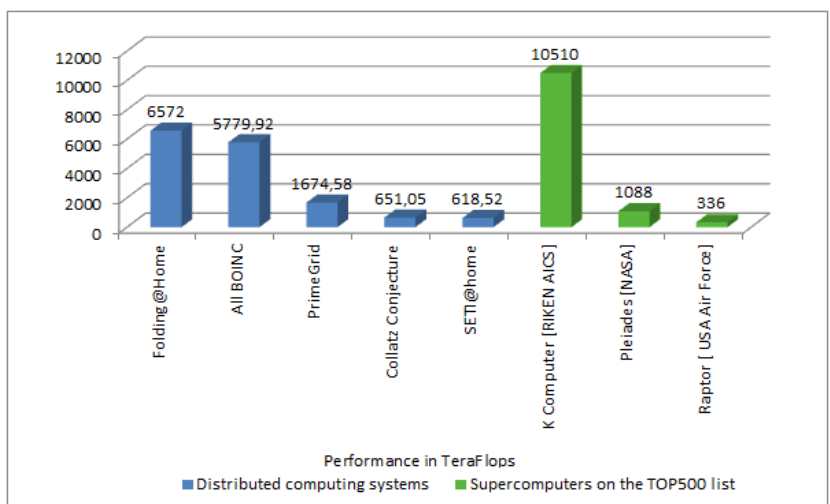
Figure 3. Performance comparison of different computational systems.

## IV. ADVANTAGES AND DISADVANTAGES OF DISTRIBUTED SYSTEMS

Distributed systems despite many advantages also have disadvantages. They are completely different in contrast to supercomputers widely adopted in the academic setting, that are typically organized in computing clusters. Distributed systems, are not able to solve all the problems, that supercomputers can handle. The most severe restriction is the data transfer rate. Distributed systems need to relay on Internet connections that are significantly slower than the solutions used in computing clusters (e.g. Infiniband interface). Therefore, tasks should not require sending large chunks of data, in fact, they should provide a long computation times on modern CPUs with relatively small input information. Another significant drawback is the calculation uncertainty. One needs to remember, that the data are being processed on third-party computers, so there is no guarantee that they will be done correctly. Moreover, there is no assurance that the computation results will be uploaded back to the server, while the user can at any time uninstall required software or reinstall their operating system.

Significant advantage of the distributed systems, is their relatively low operating cost. Within the Client/Server architecture only the server administration needs to be handled and that often means as little as one physical server. The server is utilized only to assign and manage tasks that are then being completed by other computers. Low cost of such solutions has led to high interest from amateur programmers, individual scientists and research facilities.

Fig. 3 presents a comparison of a few most important distributed systems with three supercomputers that are listed on TOP500. With regards to the overall performance, distributed systems are often not lagging far behind the technology used currently by NASA or the military (that needs significant financial capital to operate). In order to create the chart for supercomputers, LINPACK performance results that measure speed of solving a complex system of linear equations were used. On the other hand, for the distributed systems, information provided by system administrators was collected. These are most commonly based on the number of uploaded results in a given timeframe.

A simple comparison of supercomputer and a distributed system has been shown in Table I.

TABLE I. COMPARING THE CHARACTERISTICS OF A DISTRIBUTED SYSTEM AND CLASSICAL SUPERCOMPUTERS

|  | *Distributed system* | *Supercomputer* |
| --- | --- | --- |
| Reliability | Low | High |
| Independence | Low | High |
| Scalability | High | Mean |
| Cost | Very low | Very high |

There are many additional components that can help to design and build individual distributed platforms. The system can use precompiled solutions for the middleware, that facilitate effective communication between different components of the system. Most commonly used middleware platforms are CORBA (Common Object Request Broker Architecture), RMI (Remote Method Invocation) and DCOM (Distributed Component Object Model).

## V. POSSIBLE SERVER DAEMONS

Certain services need to be offered by a distributed system server i.e. task generation, distributing the tasks between the clients and analyzing returned results. Number of daemons, as well as work scheduling, do not need to be evenly distributed, however most often it does not differ significantly from the structure used by BOINC.

*a) Assimilator:* Assimilator operates on tasks that have been finished and their results are already known. This service, usually saves relevant data regarding the task to the central data base system, and when the need arise, it can also delete temporary data from the database.

*b) Transitioner:* Transitioner is responsible for analyzing tasks' status. This is the service that e.g. assigns tasks to other computers when the initial one did not return the results in a given timeframe. In the mean time the old task is cancelled. Moreover, when the same task is sent to a few different computers and different results are returned, Transitioner sends it to additional computers to verify computation correctness.

*c) Validator:* Validator is one of the last daemons handling task-related operations. This service verifies uploaded results. Generally, this daemon also assigns points (credits) to users that are considered an award for the computation contribution (this is done in the BOINC).

*d) Work Generator:* Work Generator is a daemon generating tasks in a fully automated manner, that are then distributed to other computers. These tasks can be generated considering computational capabilities of various computers (available RAM memory, free HDD space, type of CPU).

*e) Others:* There are additional daemons that can be described in the server structure. These daemons can be responsible for other operations (e.g. cleaning) on files and data base records. Due to their work, despite long operating times of the system, the amount of stored data does not increase indefinitely and waste information are not stored.

Distributed systems frequently include a browser based interface, that allows to view tasks and computer related statistics. Often, there are other services aimed to cooperate with that kind of interface.

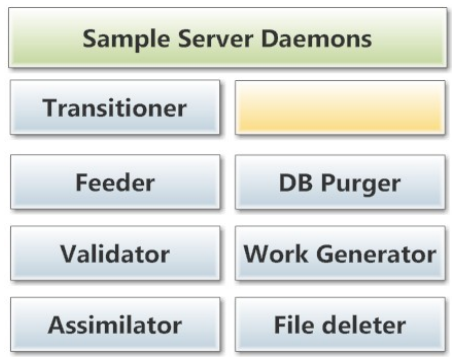

Figure 4. Separate server daemons at Berkeley Open Infrastructure for Network Computing.

## VI. SCALABILITY

Within the term scalability in distributed systems there are clearly a few key aspects that can be mentioned. Figure 5 presents these aspects.

Size scalability can pose a significant challenge. There might be a large number of users joining the system that is available on the Internet. The amount of clients and computers can not increase indefinitely, therefore at a certain point, the server will become the bottleneck. Furthermore, computers can be located in different parts of the world, sometimes even in regions where downloading larger amounts of data can result in long waiting times. Due to the large physical separation, issues related with Internet backbone can also be visible. The term scalability also defines system administration and maintenance. Despite the fact that the system is being distributed, it should be perceived by the users as one logically consistent system.

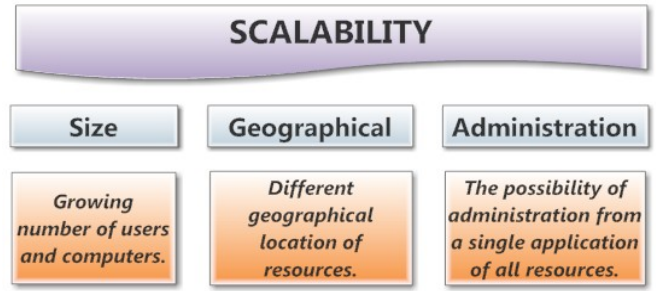

Figure 5. Main problems concerning the scalability of distributed systems.

## VII. SECURITY

Similarly to the scalability issue, during the security analysis one can denote smaller, distinct components.

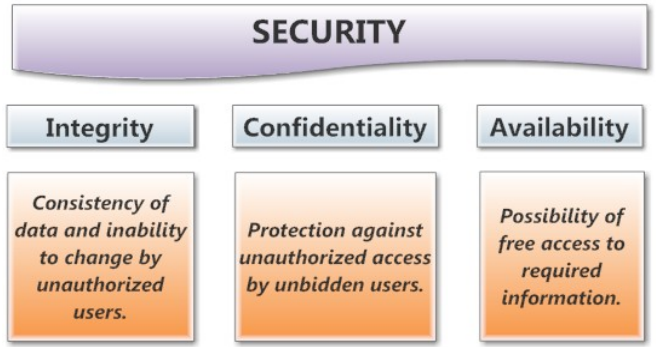

Figure 6. Main problems regarding distributed systems security

Data processed using distributed systems are more vulnerable to illegal intrusions. The platform should maximize security, however it will always be lower than with supercomputers that are operated only by certain entities.

## VIII. APPLICATIONS AND ALGORITHMS

Distributed systems have potential use not only in traditional computing tasks. Other applications using these principles are network routing, distributed databases, or ICS (Industrial Control System). Such systems are also used in aircraft flight control systems, where exchange of information between heterogeneous systems is widespread.

Algorithms executed in the high performance computing systems can be classified into the following:

*1) Parallel algorithms in shared-memory model (e.g. SMP architecture computers),*

*2)* Parallel algorithms in message-passing model *(e.g. performance clusters),*

*3)* Distributed algorithms.

The main difference between algorithms lies in the communication capabilities. In the first case, the programmer can use shared memory, which is very useful and allows to implementation a large number of problems in a simple manner (library OpenMP). In the model, using message passing interface (MPI library), the developer no longer has so much freedom, but may try to define his own logical network structure, which will be used to transfer the data. However, in the case of distributed algorithms, the designer needs to accept the fact that the network structure may be the weakest link.

## IX. GOLDBACH CONJECTURE PROJECT

In the course of this research, self-developed platform operating under BOINC was used. The goal was to verify the correctness of the Goldbach conjecture. This famous number theory problem states that every even natural number greater than 2 is the sum of two primes. For more than 250 years, no one managed to confirm or disprove this hypothesis. This problem is very well suited for distributed computing, due to easy division of tasks and complete independence of their execution. The project was supported by about 15000 computers that were connected by nearly 1000 users. The system was not promoted by any additional advertisements, which clearly shows the interest of users in such endeavors. The main problem with the BOINC server was MySQL database server, that was heavily utilized. If old results were not remove from the database for several days, storing large amount of historical data resulted in non efficient query handling by the system. Due to the available resources, only an standard hard drive was used - perhaps if an array of modern SSD hard drives were used and the tables were distributed among them the problem would not be visible. RAM is another component that is essential for a server. For use in larger projects, 4 GB of RAM is the absolute minimum. Monthly maintenance cost of a dedicated server at hosting company are approximately $50 (without activation fee). To run a distributed data processing server at home, link offering high throughput is required. Most of the connections offered for individual users may not be sufficient, because of the asymmetrical link properties (too low upload speed). There are also problems with maintaining a large number of simultaneous connections on such a link. That kind of disadvantages, can often dissuade volunteers from connecting and sharing their resources with the project. This results in an overall decrease in system performance. One of the less important server components during the Goldbach Conjecture Project was the CPU. It was not extensively utilized by any of the running services. Despite this fact, official BOINC requirements for the CPU are high e.g. for particularly demanding validators of the returned results. The average load on the CPU (Intel Celeron 1200 MHz) on a Goldbach Conjecture Project Server is shown on Fig. 7. Average system performance was approximately 10 TeraFLOPS. It has been calculated based on the number of returned task during given time frame, and the average amount of computation required to execute one of them. Performance of 10 TeraFLOPS is an astonishing result, especially when taking into account such low operating costs. Platform performance measurements are presented on Fig. 8.

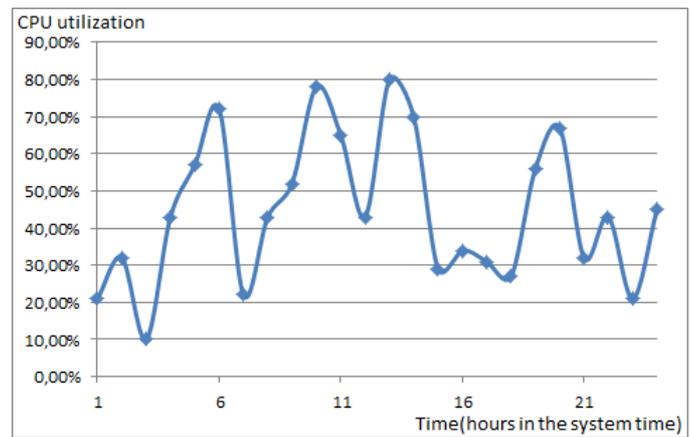

Figure 7.  Avarage server CPU utilization (Intel Celeron 1200 MHz)

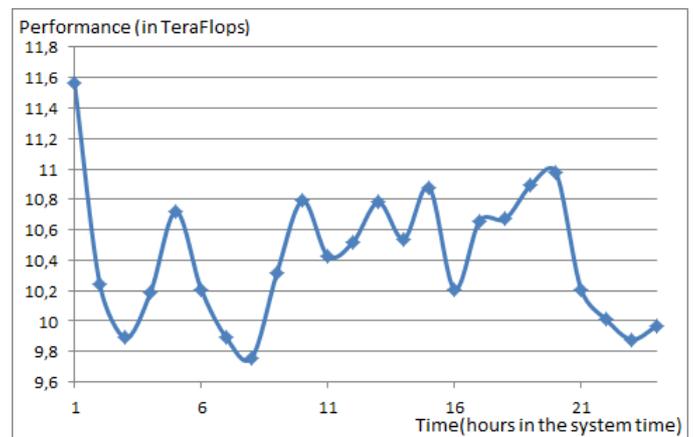

Figure 8.  Average performance platform Goldbach Conjecture Project

It is easy to realize, that performance remains fairly constant regardless of the time. Intuition suggests, that during night-time, data processing speed should drop, because many users would turn off their computers. However, one should note, that the users and their computers are scattered around the globe, and thus are located in different time zones. It is the very reason why performance level is steady. Most severe drop in performance is always caused by server problems. The whole system is dependent on its reliability. In addition, one needs to keep in mind that users participating in our project, collect points, which are stored in the database. Therefore, one needs to particularly care about not losing their achievements, since they are often the key to volunteers motivation. On average, 1.5 users per day joined the Goldbach Conjecture. From another perspective, number of computers increased every day by approximately 2. It is shown on Fig. 9 and Fig. 10 respectively. These values were measured for a system already running longer than 6 months. In case of a new BOINC project, they are initially much higher. Goldbach Conjeture Project did not solve the Goldbach's conjecture. However, working with server infrastructure developed by the University of California illustrated the great potential that resides in this solution.

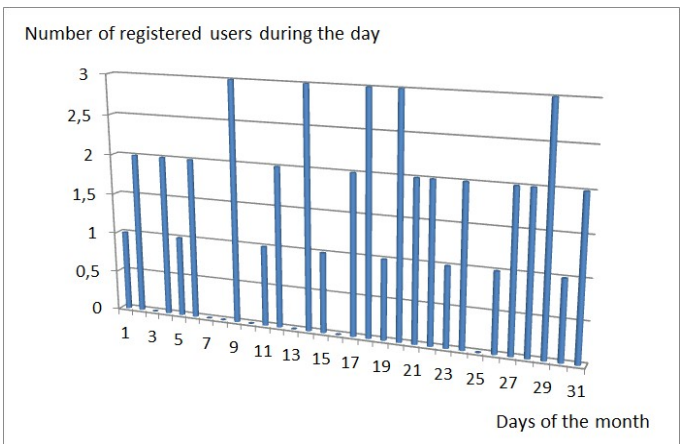

Figure 9. Number of registered users during certain days

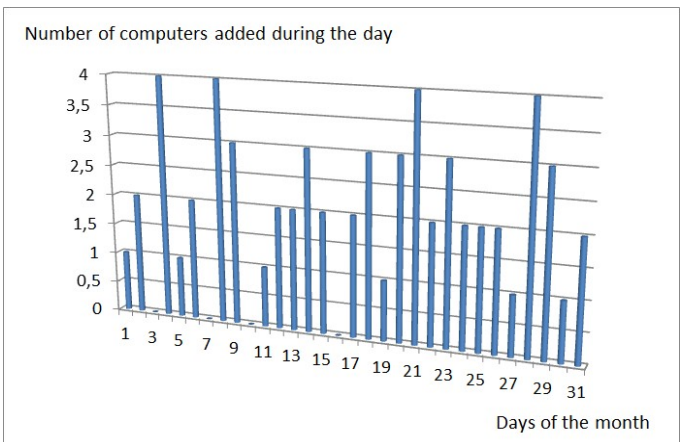

Figure 10. Number of computers added during certain days

X. CONCLUSIONS

This paper discusses the possibility of using a distributed system as a very good alternative to centralized systems, in some cases. Traditional supercomputers require significantly larger capital and operating expenditures. Nowadays, solutions based on distributed systems are becoming increasingly popular, and Internet users willingly share their computer resources to support computation carried out by academic institutions around the globe. Certainly in the coming years, this technology will grow rapidly.

However, one needs to remember that with the development of the Internet and other technologies, distributed systems will have to face new problems and challenges.